\documentclass[prl,superscriptaddress,twocolumn,preprintnumbers,%
  amsmath,amssymb]{revtex4-2}
\usepackage[utf8]{inputenc}
\usepackage{amsmath,bm}
\usepackage{graphicx,csvsimple,booktabs,epsfig}

\usepackage{color}
\allowdisplaybreaks
\usepackage[dvipsnames]{xcolor}
\usepackage[breaklinks,colorlinks=true]{hyperref}
\usepackage{mathtools}
\usepackage{amsthm,amssymb}
\usepackage{lipsum}
\usepackage{calrsfs}
\usepackage{tikz}
\usepackage{caption}
\usepackage{subcaption}
\usepackage{csvsimple}
\newcommand\myshade{85}
\colorlet{mylinkcolor}{BrickRed}
\colorlet{mycitecolor}{NavyBlue}
\colorlet{myurlcolor}{Aquamarine}
\hypersetup{
  linkcolor  = mylinkcolor!\myshade!black,
  citecolor  = mycitecolor!\myshade!black,
  urlcolor   = myurlcolor!\myshade!black,
  colorlinks = true,
}

\usepackage{url}
\newcommand*{\HH}{\mathcal{H}} 
\usepackage{xcolor}

\newcommand{\TC}{\text{TC}}  
\newcommand{\DTC}{\text{DTC}}  
\newcommand{\variat}{\partial}  
\newcommand{\abs}[1]{\vert#1\vert}

\begin{document}
\title{Gradients of O-information: low-order descriptors of high-order dependencies}

\author{T. Scagliarini}
\affiliation{Dipartimento Interateneo di Fisica, Universitá degli Studi Aldo Moro, Bari and INFN, Italy}
\author{D. Nuzzi}
\affiliation{Dipartimento Interateneo di Fisica, Universitá degli Studi Aldo Moro, Bari and INFN, Italy}
\author{Y. Antonacci}
\affiliation{Dipartimento di Ingegneria, Universitá di Palermo, Italy}
\author{L. Faes}
\affiliation{Dipartimento di Ingegneria, Universitá di Palermo, Italy}
\author{F.E. Rosas}
\affiliation{Data Science Institute, Imperial College London, United Kingdom}
\affiliation{Centre for Psychedelic Research, Department of Brain Science, Imperial College London, United Kingdom}
\affiliation{Centre for Complexity Science, Imperial College London, United Kingdom}
\affiliation{Center for Eudaimonia and Human Flourishing, University of Oxford, United Kingdom}
\affiliation{Department of Informatics, University of Sussex, United Kingdom}
\author{D. Marinazzo}
\affiliation{Department of Data Analysis, Ghent University, Belgium}
\author{S. Stramaglia}
\affiliation{Dipartimento Interateneo di Fisica, Universitá degli Studi Aldo Moro, Bari and INFN, Italy}
\affiliation{Center of Innovative Technologies for Signal Detection and Processing (TIRES), Universitá degli Studi Aldo Moro, Italy}

\begin{abstract}
    O-information is an information-theoretic metric 
    that captures the overall balance between redundant and synergistic information shared by groups of three or more variables. 
    To complement the global assessment provided by this metric, here we propose the gradients of the O-information as low-order descriptors that can characterise how high-order effects are localised across a system of interest.
    We illustrate the capabilities of the proposed framework by revealing the role of specific spins in Ising models with frustration, and on practical data analysis on 
    US macroeconomic data. 
    Our theoretical and empirical analyses demonstrate the potential of these gradients to highlight the contribution of variables 
    in forming high-order informational circuits. 
\end{abstract}
\maketitle

Network Science \cite{barabasi}, a field encompassing approaches where complex systems are represented by graphs, has grown tremendously in the last twenty years thanks to the development of powerful computational techniques to tease interdependencies out of data \cite{BOCCALETTIPR,BATTISTONPR}.
However, despite the great success of this endeavor, some important questions about complex systems cannot be properly addressed by dyadic representations, 
requiring to take into account higher-order interactions involving more than two elements. 
Such approaches typically represent 
systems as hypergraphs that can be studied via topological data analysis \cite{battiston} to reveal the structure of complex systems of interest. 

A complementary line of research focuses on emergent properties related to what the system {\it does}, and characterize its high-order behavior from observed data identifying an equivalent to hyperedges from the dynamics of the data~\cite{rosas2022}. 
A prominent role in this literature is played by the framework of partial information decomposition \cite{PID} and its subsequent developments \cite{lizier}, which exploit information-theoretic tools to evidence high-order dependencies in groups of three or more variables --- and the description of their synergistic or redundant nature. In this context, redundancy corresponds to information which can be retrieved independently from more than one source, while synergy correspond to statistical relationships that exist in the whole but cannot be seen in the parts \cite{wibral_partial_2017}: see, e.g., Ref.~\cite{luppi} for an application of these principles in neuroscience. 
Another popular computational tool is the O-information $\Omega$ \cite{rosas_quantifying_2019}, which captures the overall balance between redundant and synergistic high-order dependencies in complex systems \cite{gatica_high-order_2021,stramaglia_quantifying_2021}, whereas a positive (negative) $\Omega$ means that the multiplet of variables at hand is dominated by redundant (synergistic) dependencies. The computational complexity of $\Omega$ calculation scales more gracefully with the number of variables than PID, making it particularly well-suited for practical data analysis.

Crucially, the quest for high-order descriptions of complex systems comes with important computational and conceptual costs, as these representations often grow super-exponentially with the system size. Moreover, while coarse-grained measures such as $\Omega$ exist, their global nature fails short not being able to provide a local description of how high-order phenomena are distributed across systems of interest. 
Hence, there is a urgent need of intermediate approaches that can enable a
compact yet meaningful representation of informational multiplets. 
Indeed, the success of Network Science rests partially on the availability of metrics that quantify the role of 
specific nodes or links in the system --- metrics which are not as immediate to develop and grasp for high-order analyses, in particular when the high-order links are statistical dependencies. 

Here we address this problem by introducing a novel approach providing low-order (i.e. univariate and pairwise) descriptors of high-order dependencies in the system. 
The proposed approach is based on the gradients of the O-information: 
instead of focusing on the O-information \textit{of groups of variables}, we focus on the variation of the O-information \textit{when variables are added to the rest of the system to form these groups}. 
This provides a more nuanced descriptions of 
synergistic or redundant informational circuits, in which the role of each variable can be disambiguated. 
This new framework 
is operationalised by means of the definitions and derivations presented below.


The O-information $\Omega$ of a complex system described by $n$ stochastic variables $\bm{X}^n=(X_1,\dots,X_n)$, see \cite{rosas_quantifying_2019}, can be calculated as:
\begin{equation}
    \Omega(\bm{X}^n) =  
         (n-2)H(\bm{X}^n) + \sum_{i=1}^n \Big[ H(X_i) - H(\bm{X}^n_{-i}) \Big],
         \label{Oinfo}
\end{equation}
where $H$ is the Shannon entropy. In order to assess the contribution of a given variable $X_i$ to the informational circuit contained in $\bm X^n$, we propose to calculate its ``gradient of O-information'' given by
\begin{align}
    \variat_i\Omega (\bm{X}^n) =&\Omega(\bm{X}^n) - \Omega(\bm{X}^n_{-i})\nonumber \\=& (2-n)I(X_i;\bm{X}^n_{-i}) + \sum_{k=1,k\neq i}^{n} I(X_k;\bm{X}^n_{-ik}),
    \label{eq:grad1}
\end{align}
where $\bm{X}^n_{-i}$ denotes the set of all the variables in $\bm{X}^n$ but $X_i$, and $I$ is the mutual information~\footnote{Correspondingly, $\bm{X}^n_{-ik}$ corresponds to all the variables in $\bm X^n$ except $X_i$ and $X_k$.}. 
The quantity $\variat_i\Omega (\bm{X}^n)$ captures how much the O-information changes because of adding $X_i$, hence giving an account of how this variable contributes to the high-order properties of the system. Correspondingly, $\variat_i\Omega (\bm{X}^n)>0$ means that $X_i$ introduces mainly redundant information, while $\variat_i\Omega (\bm{X}^n)<0$ indicates that it fosters synergistic interdependencies. 

A direct calculation shows that the following bounds hold and are tight:
\begin{equation}
  -(n-2)\log|\mathcal{X}| \leq   \variat_i\Omega (\bm{X}^n)    \leq \log |\mathcal{X}|,
\end{equation}
where $|\mathcal{X}|$ is the cardinality of the largest alphabet in $\bm{X}^n$ (a proof of this can be found in the supplementary material). The asymmetry between the two bounds has an important consequence: \textit{while redundancy can be only build step by step, synergy can be established more rapidly}.

Following a similar rationale to the one that lead to Eq.~\eqref{eq:grad1}, one can further introduce a second-order descriptor of high-order interdependencies by considering gradients of gradients. In particular, the second-order gradient of a pair of variables $X_i$ and $X_j$ can be defined as
\begin{align}
    \variat_j \variat_i \Omega(\bm{X}^n) =\variat_i\Omega(\bm{X}^n) - \variat_i\Omega(\bm{X}^n_{-j}). 
\end{align}
This second-order gradient captures how much the presence of the variable $X_j$ alters the variation of O-information of the system due to the inclusion of $X_i$. It is direct to verify the symmetry $\variat_i \variat_j \Omega(\bm{X}^n) = \variat_j \variat_i \Omega(\bm{X}^n)$; therefore, we simply denote this quantity as  $\variat^2_{ij} \Omega(\bm{X}^n)$.

An interesting property of $\variat^2_{ij} \Omega(\bm{X}^n)$ is that it can be re-written as a `whole-minus-sum' property:
\begin{align}
    \variat^2_{ij} \Omega(\bm{X}^n)&= \big[ \Omega(\bm{X}^n) - \Omega(\bm{X}^n_{-ij})\big]  \nonumber\\  
    - \big[ &\Omega(\bm{X}^n_{-i}) - \Omega(\bm{X}^n_{-ij})\big] -\big[ \Omega(\bm{X}^n_{-j}) - \Omega(\bm{X}^n_{-ij})\big]~.
\end{align}
In other words, $\variat^2_{ij} \Omega(\bm{X}^n)$ measures to what degree the variation to the O-information due to the inclusions of both $X_i$ and $X_j$ is more than the sum of the variations one obtains when including them separately. Consequently, $\variat^2_{ij} \Omega(\bm{X}^n)$ vanishes if variables $X_i$ and $X_j$ are part of independent informational circuits.

The second-order gradient $\variat^2_{ij} \Omega(\bm{X}^n)$ can be compared with the {\it local O-information} between the variables $X_i$ and $X_j$ (introduced in \cite{rosas_quantifying_2019}) $I\left(X_i; X_j ; \bm{X}^n_{-ij}\right)$, which corresponds to the interaction information~\cite{mcgill} 
between  $X_i$, $X_j$ and variables $\bm{X}^n_{-ij}$. 
Interestingly, for $n=3$ the local O-information and $\variat^2_{ij} \Omega(\bm{X}^n)$ coincide, while for $n\geq 4$ they generally differ. 
A key difference between these quantities is that 
the local O-information treats the rest of the system (i.e. $\bm{X}^n_{-ij}$) as a whole, whilst in the value of the former is actually dependent on the specific partition that divides $\bm{X}^n_{-ij}$ into parts, which gives it more sensitivity to evaluate informational circuits.

Successive gradients follow automatically, resulting in a simple chain rule. If $\gamma$ is a subset of $\{1,\ldots,n\}$ of cardinality $|\gamma|$, then:
\begin{align}
                  \variat^{|\gamma|}_\gamma \Omega (\bm{X}^n)  &= \sum_{\alpha \subseteq \gamma }  (-1)^{|\alpha|} \Omega(\bm{X}^n_{-\alpha} ),
\end{align}
the sum being over all the subsets $\alpha$ of $\gamma$.
For example, for triplets of variables the gradient of the O-information reads:
\begin{align}
     \variat^3_{ijk}\Omega(\bm{X}^n)  &= \Omega(\bm{X}^n)   -   \Omega(\bm{X}^n_{-i})  -  \Omega(\bm{X}^n_{-j})  -  \Omega(\bm{X}^n_{-k})   \nonumber\\
          + & \Omega(\bm{X}^n_{-ij})  +  \Omega(\bm{X}^n_{-ik})  +  \Omega(\bm{X}^n_{-jk})  -\Omega(\bm{X}^n_{-ijk}),
\end{align}
and measures the irreducible contribution to the O-information by the triplet $\{i,j,k\}$ which cannot be ascribed to the inclusion of pairs nor single variables of the triplet. The potential of interpreting these quantities in a topological manner, as is has been done with the entropy~\cite{baudot2019topological}, is an interesting avenue for future research.

\setlength{\tabcolsep}{1pt} 

\begin{figure}
     
           \centering
    \includegraphics[width=0.95\columnwidth]{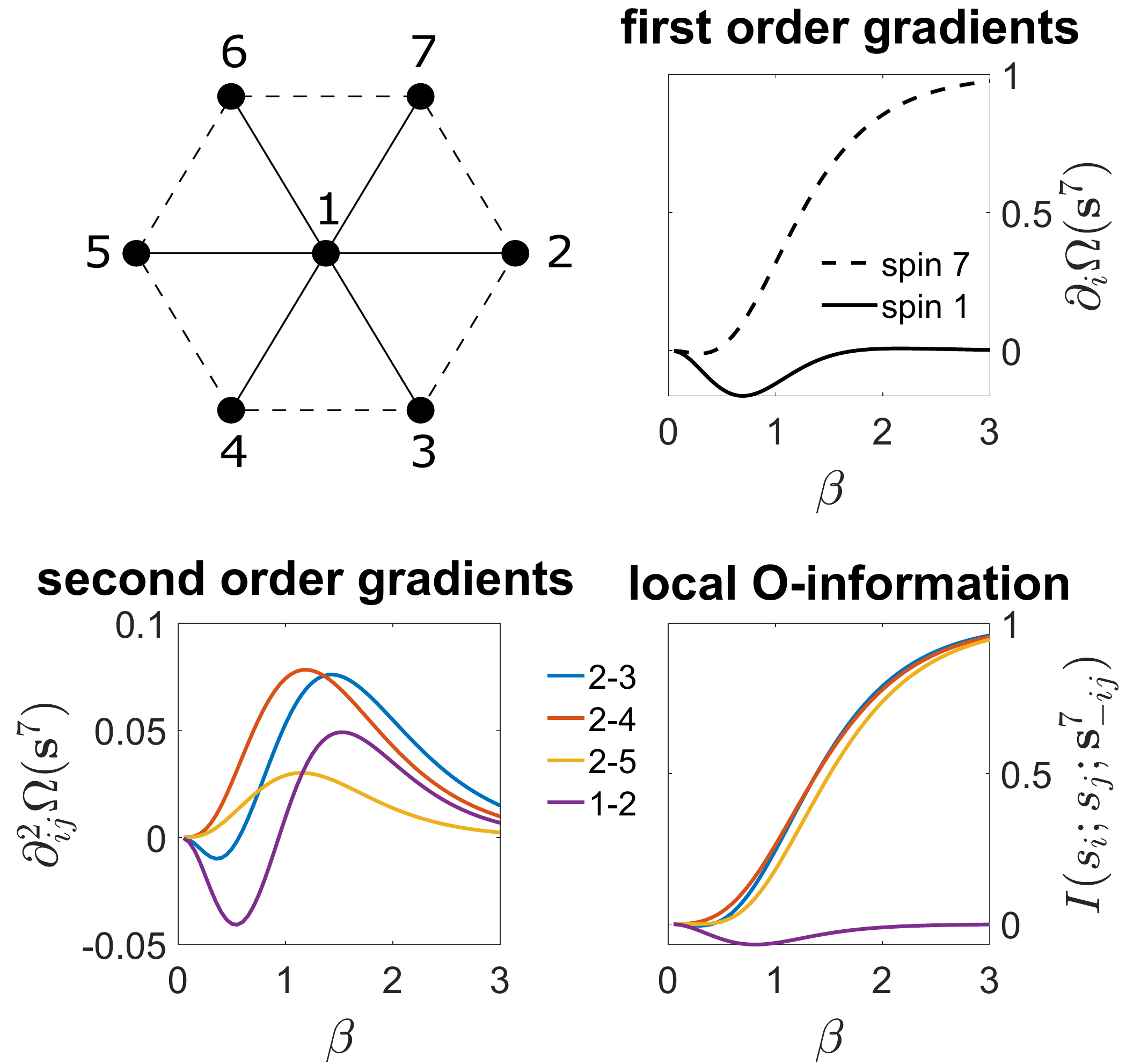} \\
     
    \caption{\textbf{Top left}: the hexagonal geometry of the Ising model, where continuous and dashed lines indicate ferromagnetic and anti-ferromagnetic interactions, respectively. \textbf{Top right}:  the gradients w.r.t. single spins $\variat_{i} \Omega(\bm{s}^7)$ are plotted versus $\beta$. For symmetry, the curves of peripheral spins 2-7 are equal. \textbf{Bottom left}: the second order gradients $\variat^2_{ij}\Omega (\bm{s}^7)$  are plotted versus $\beta$. Due to symmetry, only five non equivalent curves are plotted.  \textbf{Bottom right}: the local O-information $I(s_i;s_j;\bm{s}^7_{-ij})$ is plotted for the same pairs of spins.    \label{fig:1}}
\end{figure}

To illustrate the power of the proposed tools, let's start considering a Ising model with 
Hamiltonian given by
\begin{equation}
    \HH(\bm{s}^n) = -\sum_{i\ne j} J_{ij}s_i s_j.
\end{equation}
Our analysis considers a case where $n=7$, with couplings $J_{ij}=\pm 1$ as depicted in Fig.\ref{fig:1} (top-left panel). 
Informational measures can be computed directly from the probability of a spin configuration $\bm{s}^7 = (s_1,s_2,\ldots,s_7)$, which is given by
\begin{equation} \label{eq:prob}
    p(\bm{s}^7 ) = \frac{e^{-\beta \HH (\bm{s}^7)}}{Z}~,
\end{equation}
where $Z$ is the partition function
\begin{equation}
    Z = \sum_{s_1,\cdots,s_7}e^{-\beta \HH(\bm{s}^7)},
\end{equation}
and $\beta$ is the inverse temperature. 
%
An evaluation of $\variat{i} \Omega(\bm{s}^7)$ shows that the contribution of peripheral spins is dominated by redundancy, whilst the central spin (the one that, when added to the rest of the system, introduces frustration) introduces synergistic dependencies --- with the synergy peaking at a a finite temperature. 
These findings confirms the relationship between synergy and frustration in spin systems already noticed in \cite{musica}, while explaining which elements are most responsible for it. 
Additionally, an analysis using second-order gradients show that all the pairwise descriptors are associated with redundancy when the temperature is low, while for high temperatures 
they are synergistic for pairs consisting of the central spin and a peripheral spin, as well as for the pairs consisting of two neighboring spins on the periphery. 
Overall, these finding add an important spatial description to the previously reported~\cite{pre2019} relationship between synergy and higher temperature systems. 
We note that these findings cannot be retrieved by applications of the local O-information on the same system (Fig.\ref{fig:1}, bottom-right panel).

As an econometric application, let's consider $14$ US macroeconomic time series taken from the Federal Reserve Economic Dataset (FRED) \cite{fred_online}. We consider quarterly indicators over a period of 61 years (Apr 1959- Jan 2020) for a total of $244$ observations: 
paid compensation of employees (COE),
consumer price index  (CPIAUCSL),
effective federal funds rate  (FEDFUNDS),
government consumption expenditures and investment (GCE),
gross domestic product (GDP),
gross domestic product price deflator  (GDPDEF),
gross private domestic investment (GPDI),
ten-year treasury bond yield  (GS10),
non-farm business sector index of hours worked  (HOANBS),
M1 money supply (narrow money M1SL),
M2 money supply (broad money  M2SL),
personal consumption expenditures (PCEC),
three-month treasury bill yield  (TB3MS),
and unemployment rate  (UNRATE).
A wide literature (see e.g. \cite{smets, JUSTINIANO}) have tried leveraging similar data to address the fundamental question regarding the source of economic fluctuations. Here we are not interested in the role played by shocks and frictions neither in predicting business cycles; rather, our goal is to evidence high-order dependencies in macro-indicators of the US economy.
In order to deal with stationary time-series, the proposed approach has been applied to the logarithmic returns of the series, over which 
the gradients of the O-information were calculated using the Gaussian Copula approach described in Ref.~\cite{ince2017}. 
For each gradient, significance testing is performed via bootstrap sampling with replacement: if the  $95\%$ confidence interval of that gradient (here computed on $1000$ realizations) does not contain zero, the gradient is declared significant. Our results show that seven indicators are redundant with the rest of the system 
(see Table \ref{tab:us_indicators}) 
--- which is consistent with the prevalence of redundancy 
in real-world multivariate systems, as reflected by latent factors being typically associated to 
positive O-information (see also Ref.~\cite{marinazzo2022information}). 
In contrast, GPDI was found to play a major synergistic role with respect to the rest of the system, which may be associated with the fact that GPDI is considered a good predictor of the productive capacity of the economy.

\setlength{\tabcolsep}{8pt} 
\begin{table}[!hb]
    \centering
    \normalsize
    \begin{filecontents*}{indicatori_macroeconomici.csv}
US macroeconomics indicators, $\partial_{i}\Omega$ ,US macroeconomics indicators
COE, $0.59$,  Paid compensation of employees
HOANBS, $0.47$,  Non-farm business sector index of hours worked 
GDPDEF, $0.33$,  Gross domestic product price deflator  
UNRATE, $0.27$,  Unemployment rate  
FEDFUNDS, $0.15$,  Effective federal funds rate  
TB3MS, $0.11$,  Three-month treasury bill yield 
M2SL, $0.09$,  M2 money supply (broad money) 
GPDI, $-0.26$,  Gross private domestic investment  
 
 
 

\end{filecontents*}
\csvreader[tabular=lc, table head=\toprule, table foot=\bottomrule, no head,
    late after line=\\, late after first line=\\\midrule,]{indicatori_macroeconomici.csv}{}{\csvcoli & \csvcolii}
    \caption{Gradients of O-information for US macroeconomic indicators (only statistically significant values). }
    \label{tab:us_indicators}        
\end{table}


\begin{figure}
      \centering
    \includegraphics[width=0.95\columnwidth]{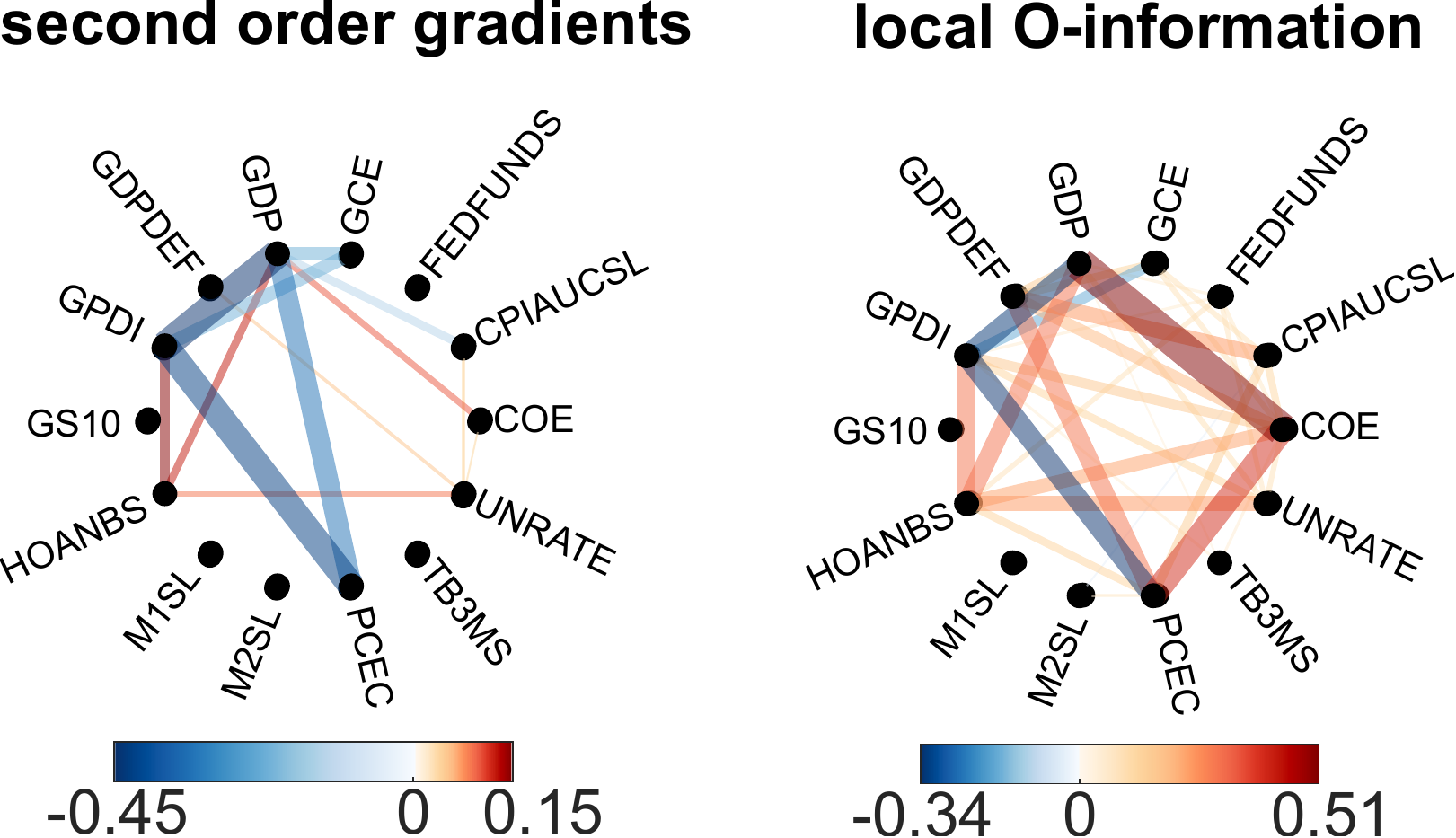} \\

    \caption{ \textbf{Left}: second order gradients for pairs of economic indicators.  \textbf{Right}: local O-information of pairs of economic indicators. Edge values are encoded by color (sign) and width (absolute value). Only statistically significant edges --- calculated via bootstrap resampling --- are included.
    } 
    \label{fig:2}
\end{figure}


These first-order analysis can be enriched by the second-order gradient. 
Results show that several pairs of variables are involved in informational circuits, as shown in Fig.~\ref{fig:2}: whilst first-order analysis shows a prevalence of redundancy, second order gradients show a prevalence of synergy. The most connected node for synergy is GDP, displaying significantly negative $\variat^2_{ij} \Omega$ with four other nodes. 
When compared with the local O-information, the proposed pairwise descriptors leads to a more sparse and parsimonuous pattern.

Summarizing, in this work we have introduced the gradients of O-information as measures of how much a variable --- or a pair of variables --- are functionally connected with the rest of a network through redundant and synergistic informational circuits. The use of this novel metric, together with measures of pairwise dependencies and global measures of higher-order information like the regular O-information, provides a more complete description of the informational character of dynamics in complex systems. By analyzing the dynamical interactions between US macroeconomics indicators we have shown how these new tools are capable of revealing high-order informational circuits, which evidenced the synergistic role of GPDI --- 
it is matter for further research to verify if existing econometric models are able to reproduce this high order behavior exhibited by data. The extension of these novel measures of high-order behavior to dynamical systems scenarios is also in order, as that
would open new avenues for investigating the function of complex networked systems on a wide range of applications.
\bibliographystyle{ieeetr}
\bibliography{biblio}
\newpage

\onecolumngrid

\begin{center}
    \textit{\LARGE Supplementary material}
\end{center}
\section{Bounds for first order gradients of the O-information}

Here we present the proof of the bounds in equation (3) in the main manuscript.
Let us consider $n$ random variables $\bm{X}^n = (X_1,X_2,\ldots,X_n)$.
Two popular extensions of the mutual information are the total correlation $\TC(\bm{X}^n)$ and the dual total correlation $\DTC(\bm{X}^n)$
\begin{equation}
    \TC(\bm{X}^n) = \sum_{i=1}^n H(X_i) - H(\bm X^n); \qquad\qquad \DTC(\bm{X}^n) = H(\bm{X}^n) - \sum_{i=1}^n H( X_i \mid \bm{X}^n_{-i}) 
\end{equation}
where $H(\cdot )$ is the Shannon entropy.
The O-information $\Omega$ of the system is given by the difference:
\begin{align}
    \Omega(\bm{X}^n) &\equiv  \TC(\bm{X}^n) -  \DTC(\bm{X}^n) =  (n-2)H(\bm{X}^n) + \sum_{k=1}^n \Big[ H(X_j) - H(\bm{X}^n_{-j}) \Big].
\end{align}
The gradient of the O-information (see the main manuscript) is given by:
\begin{align} \label{eq:dOmega1}
    \variat_i\Omega (\bm{X}^n) &= \Omega(\bm{X}^n) - \Omega(\bm{X}^n_{-i}) = (2-n)I(X_i;\bm{X}^n_{-i}) + \sum_{k=1}^{n-1} I(X_k;\bm{X}^n_{-ik}).
\end{align}
Analogously we can consider the gradient of the total correlation 
\begin{align}
        \variat_i \TC &= \TC(\bm{X}^n) - \TC(\bm{X}^n_{-i}) =  I(X_i;  \bm{X}_{-i}^n) \geq 0,
\end{align}
which, being equivalent to a mutual information, satisfies $0 \leq \variat_i\TC \leq \log|\mathcal{X}|$, where $|\mathcal{X}|$ is the cardinality of the largest alphabet in $\bm{X}^n$.
On the other hand the gradient of the dual total correlation can be written as a sum of conditional mutual information terms:
\begin{align}
    \variat_i \DTC (\bm{X}^n)&= \DTC(\bm{X}^n) - \DTC(\bm{X}^n_{-i}) \nonumber\\
    &=\sum_{k=1}^{n-1}\Big[ H(\bm{X}_{-i}^n) -H(\bm{X}^n_{-ki}) -H(\bm{X}^n) + H(\bm{X}_{-k}^n)  \Big] \nonumber\\
    &= \sum_{k=1}^{n-1} \Big[ I(X_i; \bm{X}_{-i}^n) - I(X_i; \bm{X}_{-ki}^n) \Big] \nonumber\\
    &= \sum_{k=1}^{n-1} \Big[ H(X_k\vert \bm{X}^n_{-jk}) - H(X_k\vert \bm{X}^n_{-k})  \Big] \\
    &= \sum_{k=1}^{n-1} I(X_k; X_j \vert \bm{X}^n_{-jk}) \geq 0,
\end{align}
thus implying  that $0 \leq \variat_i\DTC (\bm{X}^n)\leq (n-1)\log|\mathcal{X}|$.

Putting these results together one can find that
\begin{equation}\label{eq:bounds}
    -(n-2)\log|\mathcal{X}| \leq \variat_i \Omega (\bm{X}^n) \leq \log|\mathcal{X}|.
\end{equation}
The reminding of the proof demonstrates these bounds and their tightness.

The upper bound is trivial, indeed we have already shown that $\variat_i\DTC(\bm{X}^n) \geq 0$, hence: 
\begin{equation}
    \variat_i \Omega(\bm{X}^n) =  \variat_i \TC(\bm{X}^n) -  \variat_i \DTC(\bm{X}^n) \leq  \variat_i \TC (\bm{X})  \leq \log \abs{\mathcal{X}}.
\end{equation}

The tightness of the upper bound can be proven by showing that is achieved by the $n$-\texttt{COPY} gate, specifically by taking $X_1$ as a Bernoulli variable with $p = 1/2$ and $X_1 = X_2 = \dots = X_n$. Since we have that 
$I(X_i; \bm{X}^n_{-i}) = 1$ and $I(X_i; \bm{X}^n_{-ik}) = 1$ for $i=1,2,\ldots,n$, using Equation \eqref{eq:dOmega1} it follows that
\begin{equation}
    \variat_i\Omega(\bm{X}) = (2-n) + (n-1) = 1 ,\qquad i = 1,2,\ldots,n.
\end{equation}
This covers the case of binary random variables, but the result can be readily generalized to $\abs{\mathcal{X}} > 2$.

To prove the lower bound is a little more tricky: we start noting that $\variat_i\Omega(\bm{X}^n)$ can be written as a sum of conditional interaction information terms. Indeed it has been shown in \cite{rosas_quantifying_2019} that  O-information can be decomposed as a sum of interaction information terms
\begin{equation}
    \Omega(\bm{X}^n) = \sum_{k = 2}^{n-1} I(X_k; \bm{Y}^{k-1}_1; \bm{Y}_{k+1}^n),
\end{equation}
where we used the notation $\bm{Y}^{q}_k = (X_k, \dots, X_{q})$. 
For definiteness we fix $i = 1$, obtaining
\begin{align}
    \nonumber\variat_1\Omega(\bm{X}^n) &= \Omega(\bm{X}^n) - \Omega(\bm{X}^n_{-1}) =  \sum_{k = 2}^{n-1} I(X_k; \bm{Y}^{k-1}_1; \bm{Y}_{k+1}^n) -  \sum_{k = 3}^{n-1} I(X_k; \bm{Y}^{k-1}_2; \bm{Y}_{k+1}^n) \\\nonumber
    &= I(X_2; X_1; \bm{Y}_3^n) + \sum_{k = 3}^{n-1} \left[I(X_k; \bm{Y}^{k-1}_1; \bm{Y}_{k+1}^n) - I(X_k; \bm{Y}^{k-1}_2; \bm{Y}_{k+1}^n)\right] \\
    &= \sum_{k = 2}^{n-1}  I(X_k; X_1; \bm{Y}_{k+1}^n | \bm{Y}_2^{k-1}). \label{eq:dOmega2}
\end{align}
Now we notice that each term in the sum can be written as a difference of two conditional mutual information terms (bounded between $0$ and $\log\abs{\mathcal{X}}$), hence each term has the following bounds
\begin{equation}
  -\log\abs{\mathcal{X}} \leq  I(X_k; X_1; \bm{Y}_{k+1}^n | \bm{Y}_2^{k-1}) \leq  \log\abs{\mathcal{X}}.
  \label{eq:CondInterBounds}
\end{equation}
This implies that 
\begin{equation}
    -(n-2)\log_2\abs{\mathcal{X}} \leq \variat_1\Omega (\bm{X}^n),
\end{equation}
thus proving the lower bound.  Finally, to prove that the lower bound is tight, we consider the $n$-\texttt{XOR} gate, that is $X_1 \dots X_{n-1}$ as Bernoulli random variables with $p = 1/2$ and $X_n = (\sum_{j = 1}^{n-1} X_j ) \ \text{mod} \ 2$. Using Equation \eqref{eq:dOmega1} we have $I(X_i; \bm{X}_{-i}^n) = 1$ and $I(X_k; \bm{X}_{-ik}^n) = 0$
then 
\begin{equation}
    \variat_i\Omega = (2-n), \qquad i = 1,2,\ldots,n
\end{equation}
Then the lower bound is tight, since is achieved by the variables composing a $n$-\texttt{XOR} gate.

\section{O-information of US macroeconomic indicators: triplets and quadruplets }
\begin{figure}
    \centering
    \includegraphics[width=\linewidth]{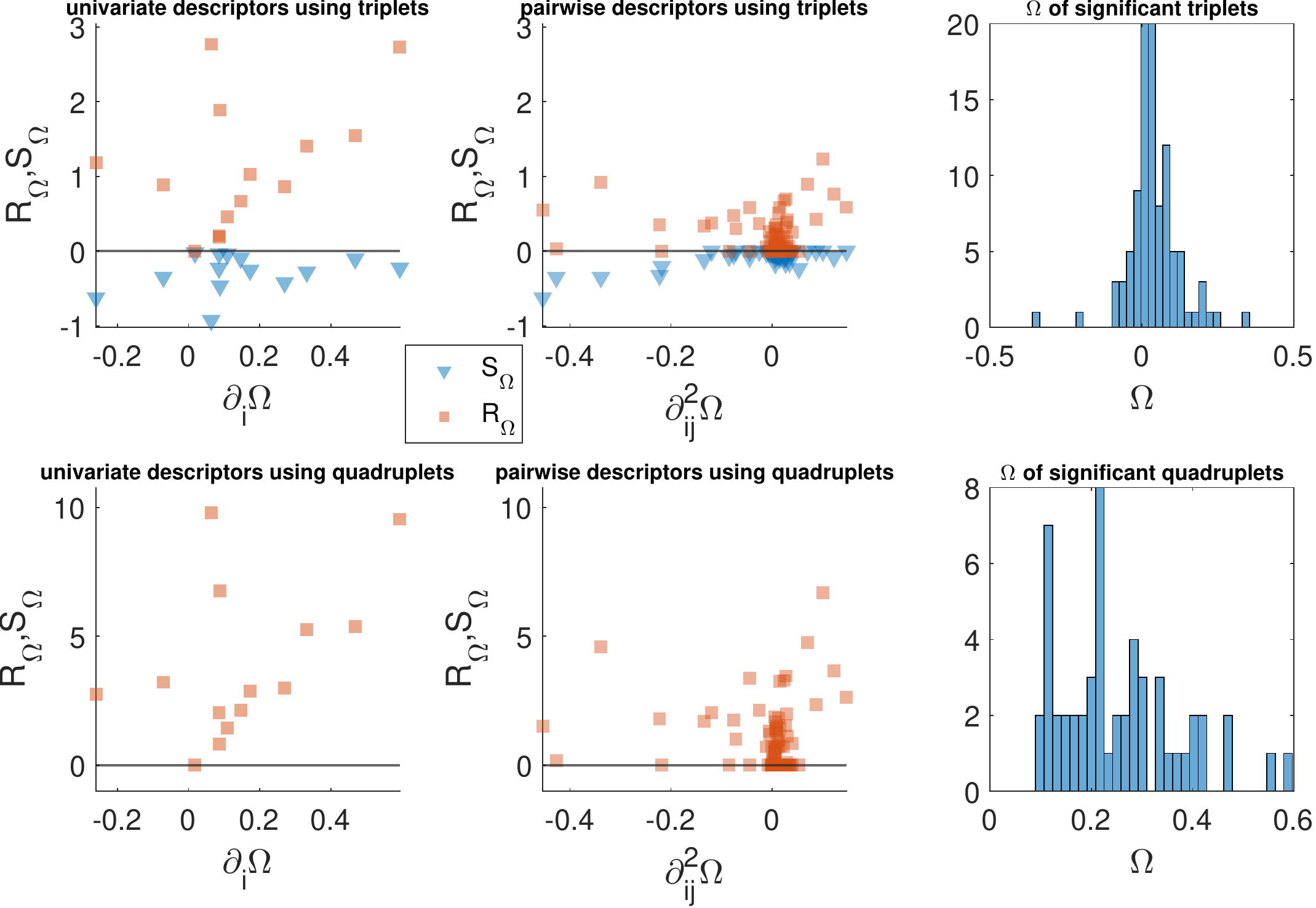}
    \caption{The first row depicts the redundancy index $R_\Omega$ and the synergy index $S_\Omega$ in the univariate  (left) and pairwise (middle) case, see the text, plotted as a function the  first order gradients and second order gradients, respectively, of the corresponding variable or pair of variables; in the right we depict the distribution of the O-information values of all the significant triplets. In the second row the same analysis has been shown for the significant O-information quadruplets. Red and blue dots indicates $R_\Omega$ and $S_\Omega$, respectively. }
    \label{fig:3}
\end{figure}

Concerning the US economic data set, we report here the conventional O-information analysis, taking into account triplets and quadruplets of variables. 
We first obtain all the triplets which are significantly synergistic and those which are significantly redundant. Then, for each variable, we sum $\Omega$ over all the redundant and significant triplets which contain that variable, obtaining $R_\Omega$ which is an index of redundancy of that variable. The same is done summing over synergistic triplets, thus leading to an index of synergy of that variable $R_\Omega$ : the results are shown in figure \ref{fig:3}top-left, where these indexes are compared with the first order gradient as found by the proposed approach. 
Analogously, for each pair of variables we sum over the triplets containing that pair, and obtain a synergy index and a redundancy index for all pairs of variables, depicted in figure\ref{fig:3}top-middle and compared with second order gradients. In figure\ref{fig:3}top-right the distribution of $\Omega$ for all the significant triplets.

In the second row of figure\ref{fig:3} the same quantities are calculated using quadruplets: we stress that no synergistic quadruplet is found to be statistically significant.

These results show that, as far as the redundancy is concerned, the proposed approach leads to a pruning of redundant pairs of variables w.r.t. the index $R_\Omega$, as shown by the vertical cloud of red points around $\variat_{ij}^2 \Omega =0$. Moreover, the redundant pattern of $R_\Omega$ is quite stable going from triplets to quadruplets. On the other hand, as far as the synergy is concerned, $\variat_{i} \Omega$ seems unrelated to $S_\Omega$ calculated on triplets.

\end{document}